\documentclass[%
 aip,
 amsmath,amssymb,
 reprint,%
]{revtex4-1}

\usepackage{graphicx}% Include figure files
\usepackage{dcolumn}% Align table columns on decimal point
\usepackage{bm}% bold math
\usepackage[utf8]{inputenc}
\usepackage[T1]{fontenc}
\usepackage{mathptmx}
\usepackage{etoolbox}

\usepackage{color}
\newcommand{\blue}{\textcolor{black}}

\newcommand{\gre}{\textcolor{black}}

\makeatletter
\def\@email#1#2{%
 \endgroup
 \patchcmd{\titleblock@produce}
  {\frontmatter@RRAPformat}
  {\frontmatter@RRAPformat{\produce@RRAP{*#1\href{mailto:#2}{#2}}}\frontmatter@RRAPformat}
  {}{}
}%
\makeatother

\begin{document}

\preprint{AIP/123-QED}

\title{Contactless Excitation of Acoustic Resonance in \blue{Insulating} Wafers}
% Force line breaks with \\

\author{Gan Zhai}
\affiliation{Department of Physics and Astronomy, Northwestern University, Evanston IL 60208, USA}
\author{Yizhou Xin}
\affiliation{Department of Physics and Astronomy, Northwestern University, Evanston IL 60208, USA}
\author{Cameron J. Kopas}
\affiliation{Rigetti Computing, Berkeley CA 94701, USA}
\author{Ella Lachman}
\affiliation{Rigetti Computing, Berkeley CA 94701, USA}
\author{Mark Field}
\affiliation{Rigetti Computing, Berkeley CA 94701, USA}
\author{Josh Y. Mutus}
\affiliation{Rigetti Computing, Berkeley CA 94701, USA}
\author{Katarina Cicak}
\affiliation{National Institute of \blue{Standards} and Technology, Boulder, CO 80305, USA}
\author{Jos\'e Aumentado}
\affiliation{National Institute of \blue{Standards} and Technology, Boulder, CO 80305, USA}
\author{Zuhawn Sung}
\affiliation{Fermi National Accelerator Laboratory, Batavia, Illinois, 60510, USA}
\author{W. P. Halperin}
\altaffiliation{w-halperin@northwestern.edu}
\affiliation{Department of Physics and Astronomy, Northwestern University, Evanston IL 60208, USA}

\date{\today}% It is always \today, today,
             %  but any date may be explicitly specified

\begin{abstract}
Contactless excitation and detection of high harmonic acoustic overtones in a thin \blue{insulator} single crystal are described using radio frequency spectroscopy techniques. Single crystal [001] silicon wafer samples were investigated, one side covered with a Nb thin film, the common starting point for fabrication of quantum devices. The coupling between electromagnetic signals and mechanical oscillation is achieved from the Lorentz force generated by an external magnetic field. This method is suitable for any sample with a metallic surface or covered with a thin metal film. High resolution measurements of the temperature dependence of the sound velocity and elastic constants of silicon are reported and compared with \blue{known results}.
\end{abstract}

\maketitle

Quantum devices essential in quantum information technology, are fabricated on silicon or sapphire wafers. In recent work it was discovered that acoustic modes in the wafer can play an important role in quantum state manipulation, including swap operations between acoustic and transmon qubit states resulting in cooling\cite{Chu.17,Chu.18}. The acoustic modes are generated from piezoelectric transducers prepared on the wafer.  This is generically  the most common method for acoustic investigation of materials, wherein electrodes are bonded to transducers that are  in direct contact with the samples of interest. The piezoelectric response of the transducer to an oscillating voltage converts the electromagnetic signal to mechanical oscillation. 
In some circumstances it is undesirable or impractical to have the electrodes or transducers in physical contact with the sample. Here we \blue{demonstrate} a contactless  technique for generating and measuring the acoustic resonances in materials.

Dobbs~\cite{Dob.73}, describes  generating acoustic resonances in metals with a solenoid and a static magnetic field. The coupling between electromagnetic signals and mechanical oscillation is achieved by the Lorentz force generated by the magnetic field, which sidesteps the use of piezoelectric materials. The Lorentz force occurs at the surface of the metal, or within the radio frequency (RF) penetration depth, generating acoustic modes in the bulk. With this approach we have studied high overtone acoustic modes  in silicon wafers and made precise measurements  of both longitudinal and transverse sound velocities and calculated the corresponding elastic constants.

Our sample is a piece of [001] single crystal silicon wafer with one side covered with a Nb thin film. The dimension of the sample cut from a commercial wafer originally 15\,cm in diameter, is  4mm\,x\,4mm\,x\,330$\mu$m (float-zone, \blue{resistivity} $>10,000\,\Omega$cm). The results described in detail in this work are for a  Nb film, 155\,nm thick, produced by \blue{High Power Impulse Magnetron Sputtering (HiPIMS)} at Rigetti Computing. High magnetic field susceptibility measurements up to 14\,T \gre{reveal that there is no temperature dependent susceptibility either before or after adding the Nb film, indicating} that there are no magnetic impurities. Additionally, we have investigated similar  samples from the National Institute of \blue{Standards and Technology (NIST)} and \gre{gold-plated} quartz transducers from Boston Piezo-Optics Inc. \gre{We find significant differences in the amplitudes and spectra of acoustic resonances of these samples as compared with what we report here which suggest that the contactless technique will be a useful characterization technique of the metal film and its interface with the wafer. The details of this continuing work are beyond the scope of the present report.}

The  radio frequency (RF) setup for contactless acoustic investigation that we report here, is the same as that commonly used for pulsed nuclear magnetic resonance (NMR) experiments. The sample is inserted in an RF coil that generates local RF fields, $H_1$, typically of order or less than 0.01\,T and placed in the magnetic field of a highly homogeneous superconducting magnet ranging from 7 to 14\,T. The RF coil provides high frequency RF pulses from the NMR spectrometer.  Reciprocally, after an excitation pulse, the acoustic resonances in the sample induce electromagnetic radiation that we detect with the same coil. This signal is processed by complex Fourier transformation to give a spectrum after minimal signal averaging for less than 10 minutes.  We report on measurements of  the spectral characteristics of the high harmonic acoustic signals under different conditions of pulsed excitation, frequency, magnetic field, and temperature from which transverse and longitudinal  sound speeds were determined and the elastic constants of single crystal silicon were determined.

\begin{figure}
	\includegraphics[scale=0.12]{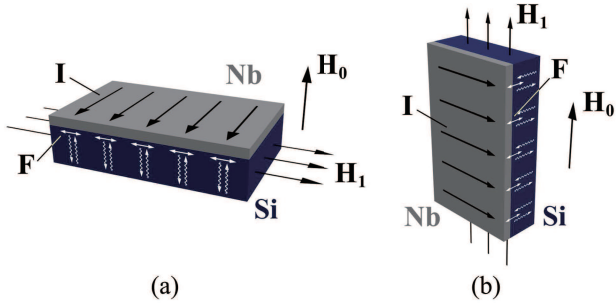}
	\caption{\label{Diagram} Schematic describing the mechanism for exciting transverse and longitudinal acoustic resonances where $H_0$, $H_1$, $I$ and $F$ represent the external magnetic field, the RF field from the pulse, the induced current, and the Lorentz force, respectively. (a) The Lorentz force is parallel to the surface, leading to a transverse mode.  (b) The Lorentz force is perpendicular to the surface, leading to a longitudinal mode.}
\end{figure}

The protocol for excitation and detection of an acoustic resonance is very similar to the procedure for single-pulse, free-induction-decay, NMR measurements with the notable exception that the direction of the RF field need not be perpendicular to the externally applied field. First, an RF pulse of a few microseconds is applied to the coil which excites an induced current sheet perpendicular to the RF field in the Nb film, depicted in Fig.~\ref{Diagram}. Since the Nb film is in an external magnetic field, the Faraday induced oscillating current in the film gives rise to an oscillating Lorentz force acting on the Nb film, causing the Nb film to oscillate producing either transverse or longitudinal acoustic modes propagating normal to the plane of the wafer. \blue{\gre{In the work we report here}, the Nb is always in the normal state.} The acoustic mode selection is determined by the relative orientation of the external field $H_0$ to that of the RF coil $H_1$. The acoustic wave is reflected from the other side of the crystal, forming an acoustic standing wave if the excitation frequency is on resonance with the natural frequency of the silicon wafer or one of its harmonics. The two orientations of the sample are shown in Fig.~\ref{Diagram} (a) and (b). For other orientations, mixed modes would be expected. 

%%%%%%%%%%%%%%%%%%%%%     Figure 1    %%%%%%%%%%%%%%%%%%%%%%%%%

\begin{figure}
	\includegraphics[scale=0.26]{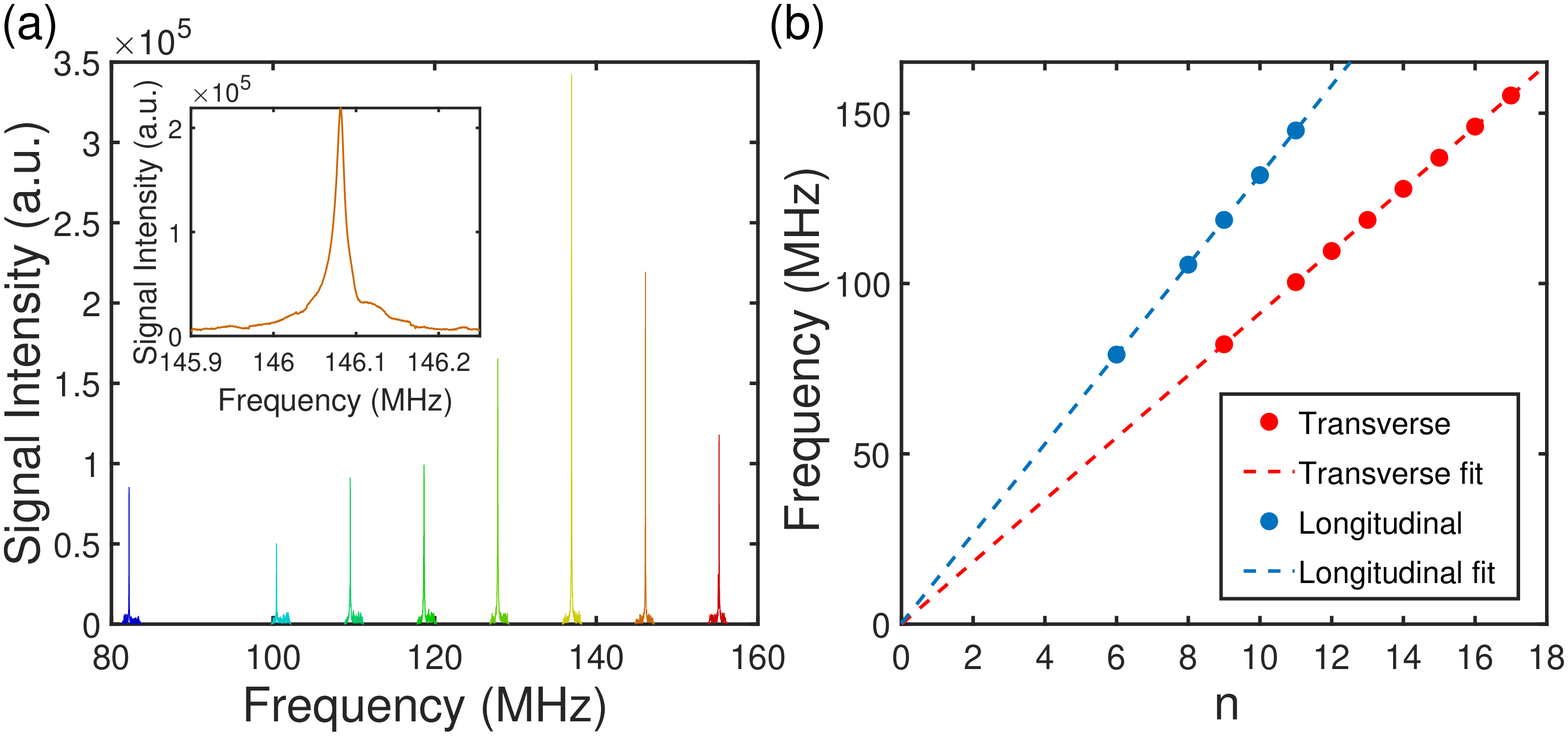}
	\caption{\label{Frequency} Harmonic frequencies of the transverse and longitudinal resonance. (a) Spectra of the transverse modes, obtained by applying frequency sweeps, each in a range of 2 MHz at constant amplitude. Inset: Magnification of the spectrum of the 16$^{th}$ harmonic. (b) Harmonic frequencies of both transverse and longitudinal modes from first moment analysis. The dashed lines are linear fits to the data. Error from the first moment of the spectra is smaller than the size of the data points.}
\end{figure}

During the detection period following the RF pulse, the induced current in the Nb film disappears leaving spectra with sharply-defined harmonic content, Fig.~\ref{Frequency}.  Induced acoustic signals observed in NMR experiments are often a nuisance\cite{Bue.78}  especially with  piezoelectric samples where the excitation is from the RF voltage.  The latter can be mitigated by inserting a metal foil within the RF coil which is not required in the present work and which we have independently shown does not effect the results we describe here.

To start our investigation, the acoustic spectra for various harmonics were measured. The n$^{th}$ harmonic frequency of the acoustic resonance in the silicon wafer is given by:

\begin{equation}
f_n=\frac{nv}{2d}
\end{equation}

\noindent where $v$ and $d$ represent the sound velocity and the thickness of the silicon wafer respectively. The acoustic signal only occurs when the resonance frequency is in the excitation bandwidth of the pulse, typically of the order of 1 MHz. A Fourier transformation of the time-domain signal following the excitation pulse gives rise to an acoustic resonance peak in the frequency domain, as shown in Fig.~\ref{Frequency} (a).  To cover a broad range of frequency the RF frequency of the pulse is shifted in steps over the desired range.  The excitation amplitude was measured independently and the tuning of the RF tuned circuit was adjusted for each step.

%%%%%%%%%%%%%%%%%%%%%     Figure 2    %%%%%%%%%%%%%%%%%%%%%%%%%

In Fig.~\ref{Frequency} (b), the data points were obtained from a first moment analysis of the spectra. The error bars of the moment analysis are smaller than the size of the data points. The two linear fit lines go to zero with very little uncertainty, confirming that the signals are acoustic in origin.

\begin{figure}[b]
	\includegraphics[scale=0.255]{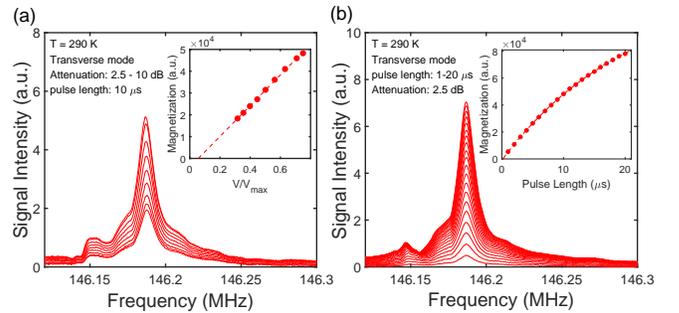}
	\caption{\label{Dependence} Signal intensity. (a) Spectra dependence on pulse voltage. Inset: signal intensity vs. pulse voltage. The dashed line represents a linear fit to the data. (b) Signal spectra under different pulse lengths. Inset: Signal intensity vs. pulse length. The dashed lines represent fits to the data.}
\end{figure}

Factors that affect the intensity of the signal were also investigated. We  found that the signal intensity mainly depends on the pulse length, the pulse voltage, and the external field,  with the pulse frequency set to the peak frequency of the spectral harmonic. The results are shown in Fig.~\ref{Dependence}. The intensities were obtained from the integral of the spectrum and the error bars are also negligible.

%%%%%%%%%%%%%%%%%%%%%     Figure 3    %%%%%%%%%%%%%%%%%%%%%%%%%

According to the calculation by Dobbs,~\cite{Dob.73} the acoustic amplitude is proportional to the RF field, as might be expected. Therefore, the signal intensity, which is proportional to the acoustic amplitude, should be proportional to the \gre{RF pulse voltage, and hence to the RF} field. The linear fit in the inset of Fig.~\ref{Dependence} (a) shows  linear dependence between the signal intensity and the pulse voltage, consistent with the theory.

From Fig.~\ref{Dependence} (b) we can see that the signal intensity has a positive correlation with the pulse length. This can be understood by the fact that for longer pulse lengths, more energy is transferred to the sample, resulting in a larger amplitude of harmonic oscillation. However, this relationship is non-linear, and it is reasonable to assume that for a long pulse length, the amplitude will reach an equilibrium value, with a characteristic time of $22.6\, \mu$s, given by an exponential fit.
%%%%%%%%%%%%%%%%%%%%%     Figure 4    %%%%%%%%%%%%%%%%%%%%%%%%%
\begin{figure}
	\includegraphics[scale=0.30]{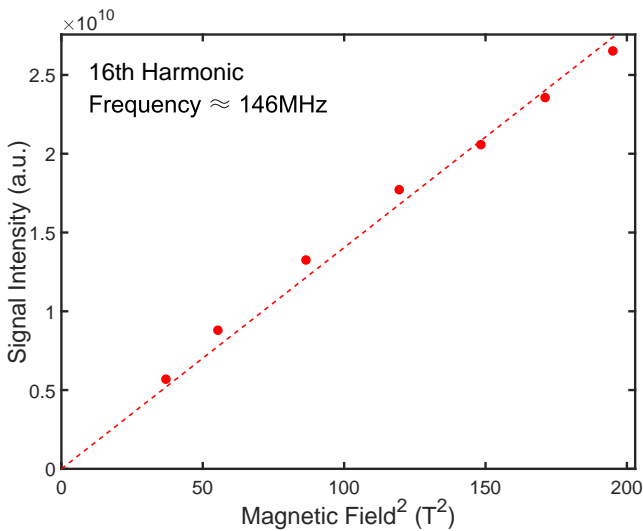}
	\caption{\label{Field} Signal intensity vs. external field squared for the transverse mode. The data points are obtained by measuring the 16$^{th}$ harmonic spectra while changing the external field. The dashed line represents a linear fit to the data \gre{forced to go through zero.}}
\end{figure}

The external field plays a role for both excitation and detection. Since the coupling strength between  electromagnetic and mechanical oscillation is proportional to the external field, we expect that the field dependence of the acoustic signal should be quadratic~\cite{Bue.78}. Our measurements in Fig.~\ref{Field} are consistent with quadratic behavior.

\begin{figure}
	\includegraphics[scale=0.3]{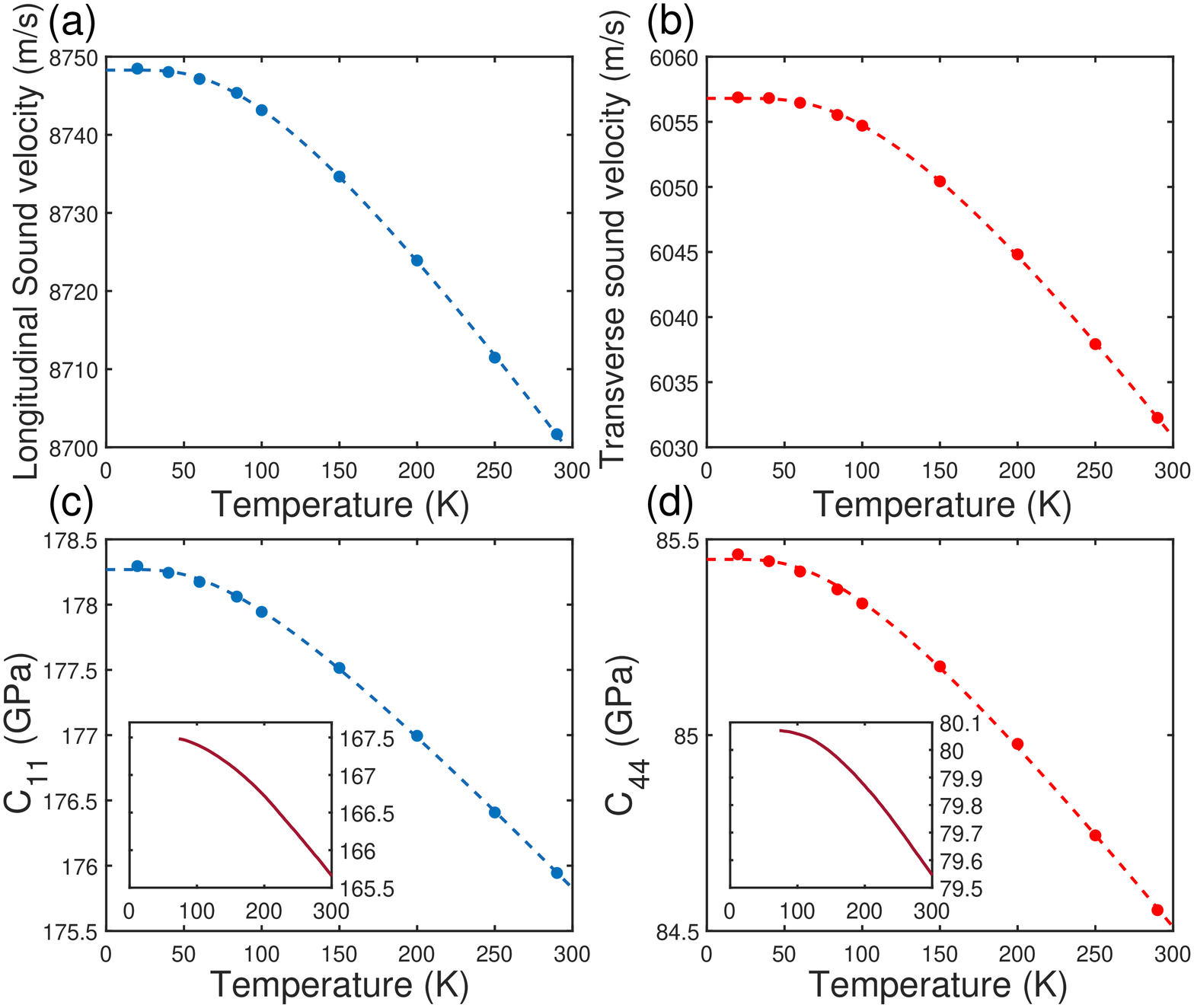}
	\caption{\label{Velocity} Temperature dependence of longitudinal (a), and transverse (b), sound velocity, and elastic constants (c, d) of the silicon wafer. The dashed lines in the four figures are fits from an empirical equation given in \blue{Ref.~5}. Inset: Measurements of $C_{11}$ and $C_{44}$ from McSkimin~\cite{McS.53} for comparison.}
\end{figure}

The technique of exciting acoustic resonance in small devices without contact has a number of potential applications. One of them is to measure the sound velocity and the elastic constants. Our measurements of the temperature dependence of sound velocities and elastic constants of a [001] silicon single crystal are shown in Fig.~\ref{Velocity}.

%%%%%%%%%%%%%%%%%%%%%     Figure 5    %%%%%%%%%%%%%%%%%%%%%%%%%

The sound velocities were obtained by measuring the temperature dependence of the harmonic frequency at a certain harmonic number. For the transverse measurement in Fig.~\ref{Velocity} we used $n =16$ and for the longitudinal measurement, $n =11$. The sound velocity was then calculated from the equation $v=2df_n/n$, where $d=330\mu\,m$ is the thickness of the silicon wafer. The temperature dependence of the thickness of the sample is taken into account in the calculation, with the data from Lyon {\it et al.}~\cite{Lyo.77}.

The elastic constants were also calculated from the sound velocities and the density of silicon. All of our measured sound velocities are for propagation in the [001] direction. Expressions for $C_{11}$ and $C_{44}$ are given in \blue{Ref.~8}.

\begin{eqnarray}
C_{11}&=&\rho v_l^2\nonumber\\
C_{44}&=&\rho v_t^2
\end{eqnarray} 

\noindent where $\rho$ is the density of silicon, and $v_l$ and $v_t$ are the longitudinal and transverse sound velocities respectively. The temperature dependence of the density is also taken into account in the calculation, with the data from Endo {\it et al.}~\cite{End.03}. The results are shown in Fig.\ref{Velocity} (c) and (d). The dashed lines in the figures are obtained by fitting the data with an empirical equation given by Wachtman {\it et al.}~\cite{Wac.61}, useful in describing the temperature dependence of elastic constants in crystals. Our results are compared with those of McSkimin~\cite{McS.53} given in the insets to Fig.\ref{Velocity} (c) and (d), who also measured the temperature dependence of $C_{11}$ and $C_{44}$ of silicon. Our measurements have the same trend as theirs. However, our values are $\approx6\%$ larger, a difference of a factor of six larger than our error of\,$1\%$\, which is  attributable to the systematic error in our measurement of the wafer thickness at room temperature.

    In summary, we have developed a technique of contactless excitation of acoustic resonances suitable for characterization of \blue{insulating} single crystals with a metallic surface, such as the silicon wafers used in quantum device applications.  This could be performed at various stages of device fabrication since the method is contactless.  We note that in \blue{Ref.~2} the alternative method of contact with a piezoelectric transducer was used to produce high harmonic acoustic signals in a wafer on which a transmon had been fabricated. At mK-temperatures \gre{ high harmonic overtones of these acoustic modes are in the quantum limit where acoustic excitations are phonon quanta}.  If stochastic excitation of acoustic modes  plays a role in limiting qubit coherence, our method can provide a useful noninvasive characterization tool. The contactless technique should also be applicable to investigate  metallic samples of quantum materials where there is coupling between magnetic degrees of freedom and high Q acoustic modes.  In this instance the RF field generates a Lorentz force limited to an RF penetration depth complementary to the acoustic modes that penetrate throughout.

The required setup for the technique is an RF circuit and an external magnetic field, similar to that used in traditional pulsed NMR experiments where excitation and detection are performed with the same coil. The frequencies of the harmonics and the thickness of the sample determine the sound velocities and  mode selection is determined by sample orientation. The detected signal intensity is proportional to the RF pulse amplitude and the square of the external magnetic field. With this technique, in the present work, the temperature dependence of both transverse and longitudinal sound velocities, as well as the elastic constants, have been precisely measured in a [001] silicon wafer where the precision is that of the frequency of the \gre{acoustic mode spectra} in the temperature range of 20\,-\,290\,K. 

\blue{This is a contribution of the National Institute of Standards and Technology, not subject to U.S. copyright. Any identification of commercial equipment, instruments, or materials in this paper is to foster understanding; it does not imply recommendation or endorsement by the National Institute of Standards and Technology, nor does it imply that the materials or equipment identified are necessarily the best available for the purpose.}

We thank Anna Grassellino, Alex Romanenko, Jens Koch, Nikolay Z. Zhelev, Man Nguyen, John Scott, and Daehan Park for useful discussion and comments.  This work was supported by the U.S. Department of Energy, Office of Science, National Quantum Information Science Research Centers, Superconducting Quantum Materials and Systems Center (SQMS) under contract No. DE-AC02-07CH11359.

%\section*{Data Availability Statement}

%The data that support the findings of this study are available from the corresponding author upon reasonable request.

\bibliography{aipsamp}% Produces the bibliography via BibTeX.

\end{document}